\documentclass[final,5p,times,twocolumn,authoryear]{elsarticle}

\usepackage{amssymb}
\usepackage{algorithm2e} 
\usepackage{graphicx}
\usepackage{multirow} 

\usepackage{amsmath}
\usepackage{hyperref}
\usepackage{orcidlink}
\usepackage[utf8]{inputenc}
\usepackage{geometry}
\usepackage{natbib} 
\usepackage{xcolor} 
\usepackage{array}

\usepackage{caption}
\usepackage{listings}
\usepackage{float}

\usepackage{booktabs}
\usepackage{tabularx}

\usepackage[T1]{fontenc}

\usepackage{tikz}
\usetikzlibrary{positioning, arrows.meta, calc}

\journal{Astronomy $\&$ Computing}

\begin{document}

\begin{frontmatter}


\title{A High-Throughput AES-GCM Implementation on GPUs for Secure, Policy-Based Access to Massive Astronomical Catalogs}

\author[inst1,inst2,inst3]{Samuel Lemes-Perera\orcidlink{0000-0003-1044-154X}\corref{cor1}}
\ead{samuel@lightbridges.es}

\author[inst1,inst4,inst5]{Miguel R. Alarcon\orcidlink{0000-0002-8134-2592}}
\author[inst2]{Pino Caballero-Gil\orcidlink{0000-0002-0859-5876}}
\author[inst1,inst4,inst5]{Miquel Serra-Ricart\orcidlink{0000-0002-2394-0711}}

\cortext[cor1]{Corresponding author}
\affiliation[inst1]{organization={Light Bridges S.L.},
addressline={Observatorio Astronómico del Teide, Carretera del Observatorio del Teide, s/n},
city={Santa Cruz de Tenerife},
postcode={E-38500},
country={Spain}}

\affiliation[inst2]{organization={Departamento de Ingeniería Informática y de Sistemas, Universidad de La Laguna (ULL)},
addressline={Calle Padre Herrera s/n},
city={San Cristóbal de La Laguna},
postcode={E-38206},
country={Spain}}

\affiliation[inst3]{organization={Instituto Tecnológico y de Energías Renovables (ITER)},
addressline={Polígono Industrial de Granadilla s/n},
city={Granadilla de Abona},
postcode={E-38600},
country={Spain}}

\affiliation[inst4]{organization={Instituto de Astrofísica de Canarias (IAC)},
addressline={C/ Vía Láctea s/n},
city={San Cristóbal de La Laguna},
postcode={E-38205},
country={Spain}}

\affiliation[inst5]{organization={Departamento de Astrofísica, Universidad de La Laguna (ULL)},
addressline={Av. Astrofísico Francisco Sánchez s/n},
city={San Cristóbal de La Laguna},
postcode={E-38206},
country={Spain}}


\begin{abstract}

The era of large astronomical surveys generates massive image catalogs requiring efficient and secure access, particularly during pre-publication periods where data confidentiality and integrity are paramount. While Findable, Accessible, Interoperable, and Reusable (FAIR) principles guide the eventual public dissemination of data, traditional security methods for restricted phases often lack granularity or incur prohibitive performance penalties. To address this, we present a framework that integrates a flexible policy engine for fine-grained access control with a novel GPU-accelerated implementation of the AES-GCM authenticated encryption protocol.

The novelty of this work lies in the adaptation and optimization of a parallel tree-reduction strategy to overcome the main performance bottleneck in authenticated encryption on GPUs: the inherently sequential Galois/Counter Mode (GCM) authentication hash (GHASH). We present both the algorithmic adaptation and its efficient execution on GPU architectures. Building on optimized GPU AES kernels from recent work in cryptographic acceleration, this work presents the first integration of these techniques into a high-throughput, FITS-aware encryption framework specifically designed for large-scale astronomical data, combining cryptographic authentication, dual-key access control, and direct compatibility with the standard astronomical Python ecosystem. Our implementation transforms the sequential GHASH computation into a highly parallelizable, logarithmic-time process, achieving authenticated encryption throughput suitable for petabyte-scale image analysis.

Our solution provides a robust mechanism for data providers to enforce access policies, ensuring both confidentiality and integrity without hindering research workflows, thereby facilitating a secure and managed transition of data to public, FAIR archives.
\end{abstract}


\end{frontmatter}


\section{Introduction}
\label{sec:introduction}

Modern astronomy, driven by large-scale surveys such as the Legacy Survey of Space and Time (LSST), is generating petabyte-scale image catalogs \citep{ivezic_lsst_2019,berriman_how_2011}. While these massive datasets are the engine of discovery, they introduce critical challenges in data security and access control, particularly during the embargo and validation phases that necessarily precede public release. Protecting sensitive data during proprietary periods, enforcing granular collaboration policies, and ensuring the priority of discoveries are paramount requirements before data fully adheres to the Findable, Accessible, Interoperable, and Reusable (FAIR) principles in public archives \citep{wilkinson_fair_2016, chen_best_2022}.

The risks associated with inadequate data protection are vividly illustrated by historical incidents ranging from ethical breaches to infrastructure attacks. The controversy surrounding the discovery of the dwarf planet Haumea \citep{brown_discovery_2005,brown_collisional_2007} highlighted how unauthorized access to observation logs can lead to priority disputes, undermining scientific integrity. More recently, the vulnerability of astronomical infrastructure was underscored by the 2022 cyberattack on the Atacama Large Millimeter/submillimeter Array (ALMA), which forced a suspension of scientific operations and data delivery, demonstrating that major facilities are targets for direct cyber threats \citep{national_radio_astronomy_observatory_alma_2022}. Similarly, the risks of utilizing unvalidated, preliminary data products were highlighted by the BICEP2/Planck controversy \citep{ade_detection_2014,ade_joint_2015}.

Beyond preventing malicious incidents, robust access control is a standard operational requirement for major facilities. Missions and observatories such as the James Webb Space Telescope, Euclid, ALMA, and LSST enforce specific data rights policies that include proprietary periods ranging from months to years. During these phases, data access must be strictly limited to Principal Investigators or consortium members for calibration and initial scientific exploitation. Furthermore, modern large-scale collaborations require granular permission management, distinguishing between core team members, external collaborators, and students, while simultaneously guaranteeing the integrity and authenticity of the data against corruption or tampering.

However, the imperative to secure these datasets often conflicts with the high-throughput access demanded by modern scientific workflows. Traditional cryptographic measures can impose significant computational overhead (latency and reduced bandwidth), creating unacceptable bottlenecks for data-intensive analysis. Addressing this dual challenge requires solutions that provide robust, fine-grained access control without compromising the I/O performance required by High-Performance Computing (HPC) environments.

To address this, we present a framework that integrates a flexible policy engine for access control with a novel, high-throughput GPU-accelerated implementation of the AES-GCM (Galois/Counter Mode) protocol \citep{dworkin_recommendation_2007, salowey_aes_2008}. Securing high-resolution astronomical images presents unique challenges: the sheer volume of pixel data requires algorithms that can sustain multi-gigabit throughput to avoid I/O starvation, while the structured nature of FITS files demands mechanisms to prevent pattern leakage. AES-GCM is widely recommended for such high-bandwidth applications because it combines the parallelizability of Counter Mode, essential for handling large payloads efficiently, with a built-in integrity check that detects data corruption without the overhead of a separate HMAC pass \citep{mcgrew_galoiscounter_2004, an_highly_2020}. Unlike simpler modes, such as Counter (CTR) that provide only \emph{confidentiality}, AES-GCM provides Authenticated Encryption with Associated Data (AEAD), ensuring both confidentiality and data integrity.

The primary technical challenge of this approach, and the core contribution of our work, is overcoming the performance bottleneck inherent in AES-GCM's authentication component: the GHASH function. GHASH operates by chaining multiplications over a binary Galois Field ($GF(2^{128})$) to compute a message authentication code. While mathematically robust, this chain implies that the processing of each data block strictly depends on the result of the previous one. Consequently, the standard GHASH algorithm is inherently sequential, making it ill-suited for the massive parallelism of Graphics Processing Units (GPUs) \citep{manavski_cuda_2007, lee_parallel_2025}. Following the strategy described by \citet{lee_parallel_2025}, we transform the linear $O(N)$ GHASH computation into a scalable, logarithmic-time $O(\log N)$ process. This enables authenticated encryption at throughputs matching the I/O capabilities of modern storage systems.

Our framework assumes the existence of a secure, external Key Management Service (KMS) for key provisioning, focusing strictly on the cryptographic performance and policy enforcement layer. This solution offers a robust mechanism for data providers to enforce access policies during sensitive periods, ensuring both confidentiality and integrity without hindering research workflows, thereby facilitating a secure and managed transition of data to public, FAIR archives. 

Building on prior work in optimized GPU AES implementations \citep{lee_speed_2024, lee_parallel_2025, tezcan_optimization_2021}, \texttt{GpuFitsCrypt} introduces a FITS-compatible authenticated container (Section~\ref{sec:architecture}) that is syntactically valid FITS, opens cleanly in Astropy and cfitsio without modification, and provides both cryptographic authentication through AES-GCM authentication tags and accidental-corruption detection through the FITS-standard CHECKSUM/DATASUM mechanism. The encrypted pixel payload is encoded as a single-row, single-column BinTableHDU regardless of the original image dimensions, concealing geometry from unauthorized accessors while preserving scientific invariance. A dual-key access policy mechanism manages Header Keys and Data Keys independently, enabling four distinct access scenarios that decouple metadata visibility from data confidentiality and support embargo-period workflows (Section~\ref{sec:secure_access}, Table~\ref{tab:policy_examples}). Critical metadata is cryptographically bound to the ciphertext via AES-GCM's Associated Authenticated Data field, preventing context-swapping attacks; the framework further enforces verify-then-decrypt semantics, rejecting authentication failures before any plaintext is exposed. Python bindings integrate \texttt{GpuFitsCrypt} with Astropy's \texttt{open()} workflow, allowing users to access encrypted FITS files with the same idiomatic patterns used for unencrypted files when valid keys are available (Table~\ref{tab:api}).

The remainder of this paper describes the data context and access challenges (Section~\ref{sec:data_challenges}), details the framework's architecture (Section~\ref{sec:architecture}), its mechanisms for secure access (Section~\ref{sec:secure_access}), the specifics of our GPU-accelerated implementation and its performance benchmarks (Section~\ref{sec:implementation_performance}), followed by a discussion (Section~\ref{sec:discussion}) and some conclusions (Section~\ref{sec:conclusions}).

\section{Data Context and Access Challenges}
\label{sec:data_challenges}

The framework operates within the specific context of modern astronomical data archives, particularly those dominated by large image catalogs. Understanding the nature of this data and the typical ways astronomers interact with it is crucial for appreciating the security and performance challenges involved.

\subsection{Characteristics of Astronomical Image Catalogs}

The vast majority of astronomical image data, from ground-based telescopes to space missions, adheres to the Flexible Image Transport System (FITS) standard \citep{greisen_fits_1980,wells_fits_1981, ponz_fits_1994, iau_fits_working_group_definition_2016}. FITS provides a crucial structure based on Header/Data Units (HDUs), recognizing the distinct yet complementary roles of metadata and pixel arrays in scientific analysis. Typically, a FITS file contains a primary HDU holding the main image array (the core scientific data, e.g., pixel values) and its associated descriptive header. Additional extensions can follow, containing other data arrays (like weight maps or masks) or binary tables, each with their own header.

Critically, the header sections store essential metadata as keyword/value pairs. This includes observational parameters (telescope telemetry, instrument configuration, exposure time, filters), processing history (calibration steps, software versions), and arguably most importantly, the World Coordinate System (WCS) information \citep{greisen_representations_2002, calabretta_representations_2002} and flux calibration (Analog-to-Digital Units, ADU, to magnitudes). The WCS maps pixel coordinates to celestial coordinates (e.g., Right Ascension and Declination), while flux calibration enables the conversion of pixel values to physical units. This rich metadata is often as scientifically valuable as the pixel data itself, enabling data discovery and context, thus necessitating careful consideration in any security framework regarding how both components are accessed and protected.

The scale of data generation in modern astronomy is expanding rapidly across all tiers of observation. This trend is evident in agile, high-cadence robotic facilities such as the Two-meter Twin Telescope (TTT)\footnote{\url{https://ttt.iac.es}} and the Transient Survey Telescope (TST)\footnote{\url{https://tst.iac.es}}, as well as in massive synoptic surveys like LSST \citep{ivezic_lsst_2019}, Pan-STARRS \citep{chambers_pan-starrs1_2016}, ZTF \citep{bellm_zwicky_2019, yao_ztf_2019}, and space missions like Euclid \citep{laureijs_euclid_2011}. Collectively, these projects generate enormous quantities of FITS data, where individual images can range from megabytes to gigabytes. Archives routinely host millions of such files, pushing total data volumes into the petabyte and exabyte scales \citep{berriman_how_2011}. Managing, querying, and accessing this data efficiently is a significant challenge that is further complicated when strict security requirements are applied.

\subsection{Common Data Access Patterns and Security Integration}

Astronomers interact with image archives in various ways, each presenting different implications for integrating security measures:

\begin{itemize}
\item Full File Retrieval: Users often download entire FITS files for local processing. For security, this implies decrypting the complete file, requiring efficient throughput handling for potentially large files (e.g., $>$500MB).
\item Metadata Queries: Researchers frequently search and retrieve information solely from FITS headers (e.g., finding all observations of a specific object or within a certain time range) without immediately needing the pixel data. Security policies might allow broader access to metadata while restricting pixel data access. The framework must therefore be able to differentiate between HDU types, allowing performant decryption of headers without the overhead of decrypting the full data payload.
\item Direct/Streaming Analysis: Modern workflows increasingly utilize libraries (e.g., Astropy \citep{collaboration_astropy_2013, collaboration_astropy_2018}) or server-side platforms to access and analyze data arrays directly, potentially without creating intermediate files on disk. The security layer must integrate transparently with these tools, performing decryption on-the-fly as data is read or streamed into memory.
\end{itemize}

A successful security framework must accommodate these diverse patterns, providing protection without unduly hindering common scientific workflows or introducing prohibitive performance penalties, especially for interactive analysis or large-scale processing pipelines.

\subsection{Limitations of Existing Security Approaches and the Role of FAIR Principles}

Current security practices in many astronomical archives often fall short of providing the required granularity, performance, and transparency for data during restricted phases:

\begin{itemize}
\item Perimeter Security (VPNs, Firewalls): While useful for network-level protection, these offer coarse-grained ("all-or-nothing") access once a user is authenticated within the perimeter. They do not provide fine-grained control over specific datasets or enforce complex policies, nor do they protect data once it has been downloaded or moved.
\item Web Portal Authentication: Basic username/password access to data portals often lacks granular authorization based on data properties or user roles beyond the initial login. Furthermore, data downloaded through such portals is typically unprotected thereafter ("data at rest").
\item Operating System Filesystem Permissions: Standard Portable Operating System Interface permissions are insufficiently expressive to capture complex scientific access policies (e.g., time-based embargoes, role-based access tied to collaboration status). Managing these permissions across petabyte-scale archives with millions of files is administratively unscalable.
\item Lack of Widespread Encryption: Due to perceived performance overheads and the complexity of key management, end-to-end encryption of large scientific data archives is not yet common practice. Data is often stored unencrypted, leaving it vulnerable if storage systems are compromised or physical media is lost.
\item Alignment with Open Science Goals: While the community strongly moves towards Open Science and FAIR data principles \citep{wilkinson_fair_2016, chen_best_2022}, facilitated by public archives \citep{green_report_2024}, these principles primarily apply once data enters the public domain. Existing mechanisms often lack dedicated solutions to robustly manage the \emph{transition phases}, embargo periods, collaborative analysis, pre-publication review, where controlled, secure access is temporarily necessary before full public release.
\end{itemize}

These limitations highlight the need for a dedicated framework that bridges the gap, integrating strong, policy-driven access control directly with performant cryptographic protection at the data level. This ensures security during the research phase while facilitating a smooth eventual transition towards public, FAIR accessibility.

\section{Framework Architecture}
\label{sec:architecture}

\begin{figure*}[t] 
 \centering
 \includegraphics[width=0.75\textwidth]{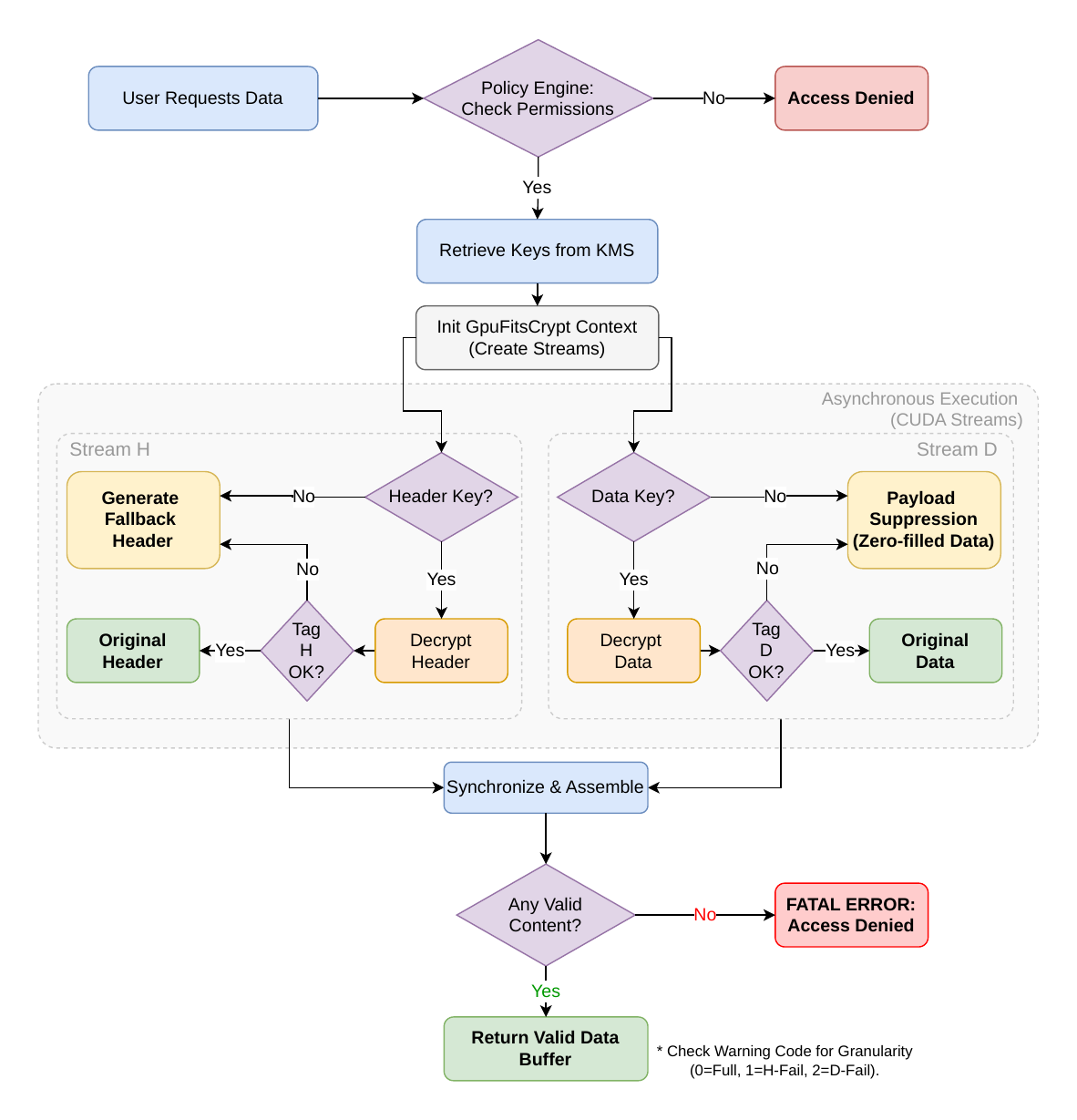} 
 \caption{Architecture and granular enforcement workflow. 
 The diagram integrates the Control Plane components with the parallel Data Plane execution. 
 Unlike monolithic approaches, it illustrates the compartmentalized verification logic: 
 specific keys determine the execution path in independent streams (H/D). 
 Crucially, the workflow demonstrates the fail-safe mechanisms: absence of a Data Key or integrity failure triggers Payload Suppression (zero-filling), while Header issues activate Fallback Header generation, ensuring a valid FITS container is returned even in partial access scenarios.}
 \label{fig:policy_workflow_diagram}
\end{figure*}

To address the dual challenges of security and performance in astronomical archives, a modular framework designed for seamless integration with existing data centers is proposed. The architecture strictly separates the concerns of policy enforcement (Control Plane) from high-performance cryptographic execution (Data Plane). Figure~\ref{fig:policy_workflow_diagram} illustrates the complete system architecture, detailing the interaction flow between the conceptual components and the parallel cryptographic streams. While this paper focuses on the implementation and performance of the GPU-accelerated cryptographic library, the architectural context is essential to understand how granular access is enforced.




The data access workflow follows a strict ``verify-then-decrypt'' orchestration. An application's request is intercepted by an Access Control Module (ACM), which validates the user's identity and queries a Policy Engine. Upon a GRANT decision, the ACM retrieves the specific decryption keys from an external KMS and passes them to the Cryptographic Module. This module then executes the high-throughput, GPU-accelerated decryption and integrity verification, returning the plaintext FITS data to the application only if both the identity authorization and the cryptographic signature verification succeed.

\subsection{Conceptual Components: The Control Plane}

The framework relies on three conceptual components to manage authorization and key distribution, ensuring that cryptographic operations are policy-driven rather than ad-hoc.

\begin{itemize}
\item Policy Engine: This module acts as the authorization brain, storing and evaluating granular access rules. Policies link user identities (e.g., Co-Investigators, Students) to data attributes (e.g., \texttt{release\_date}, \texttt{proposal\_id}) using logic such as \texttt{allow access if (user.role == 'collaborator' AND data.embargo\_active == true)}. It returns a binary \texttt{GRANT} or \texttt{DENY} decision.
\item Access Control Module: The ACM serves as the Application Programming Interface (API) gateway and orchestrator. It handles authentication, requests policy decisions, and interfaces with the KMS. Crucially, it isolates the keys from the end-user; the user never sees the raw key, only the resulting data stream.
\item Key Management Service Interface: The architecture assumes the existence of a secure, industry-standard KMS (e.g., HashiCorp Vault, AWS KMS), accessed over a TLS/mTLS-secured channel for all ACM-to-KMS interactions. The framework abstracts key generation and lifecycle management, interacting with the KMS strictly to retrieve keys for authorized sessions.
\end{itemize}

\subsection{The Cryptographic Module: \texttt{GpuFitsCrypt}}

The core technical contribution of this work is the implementation of the data plane: a high-performance C++/CUDA library named \texttt{GpuFitsCrypt}. This library handles the computationally intensive tasks of encryption, decryption, and authentication.

\subsubsection{Cryptographic Protocol: AES-GCM}
The framework implements the AES-128-GCM standard \citep{dworkin_recommendation_2007}. Unlike the CTR mode used in early prototypes which provided only confidentiality, GCM is an AEAD mode. It generates a 128-bit Authentication Tag that guarantees both the confidentiality of the pixel data and the integrity of the file. This ensures that any unauthorized modification to the encrypted payload or the metadata is immediately detected (resulting in a decryption failure), a critical requirement for scientific data provenance and validation.

\subsubsection{Authenticated FITS File Format}
To support granular access while maintaining the integrity of the file structure, a FITS-compatible authenticated container is employed (see Figure~\ref{fig:fits_format}). The original FITS file is transformed into a two-HDU structure:

\begin{figure}[htbp]
    \centering
    \includegraphics[width=\columnwidth]{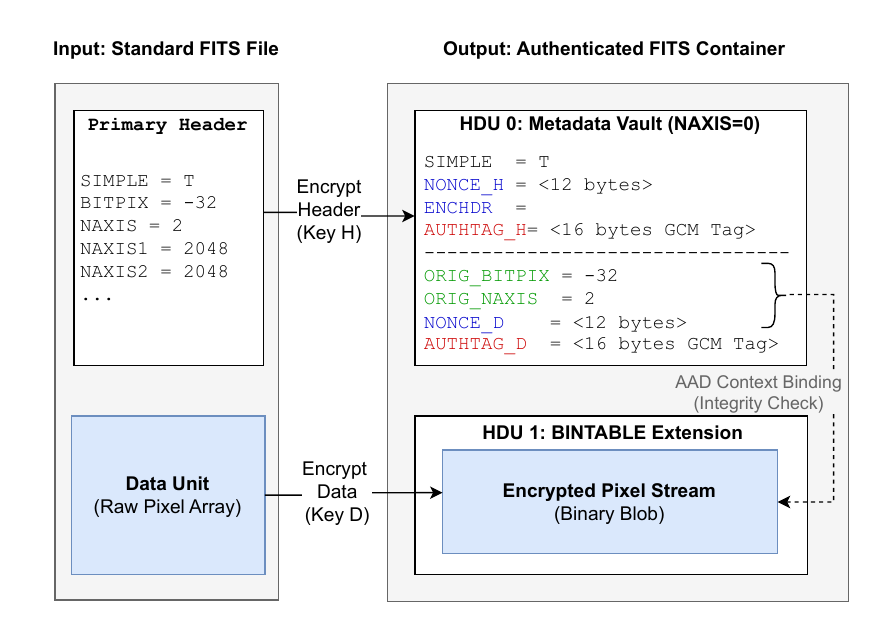}
    \caption{\textbf{Structure of the Authenticated FITS Format.} The original file (left) is split into a dual-key container (right). The Primary HDU (HDU 0) acts as a metadata vault, storing cryptographic nonces, authentication tags, and the original header encrypted within the \texttt{ENCHDR} keyword. Critical structural metadata (\texttt{ORIG\_*}) remains in plaintext to allow file parsing but is cryptographically bound to the encrypted pixel data (HDU 1) via the AES-GCM Associated Data (AAD) mechanism. This ensures that any tampering with the plaintext WCS or dimensions triggers an authentication failure in \texttt{AUTHTAG\_D}.}
    \label{fig:fits_format}

\end{figure}

\begin{enumerate}
\item Primary HDU (Metadata Container): A metadata-only HDU (\texttt{NAXIS=0}) acting as a secure envelope containing:
\begin{itemize}
\item \texttt{NONCE\_H} and \texttt{NONCE\_D}: 96-bit (12-byte) initialization vectors generated via a Cryptographically Secure Pseudo-Random Number Generator. These ensure uniqueness for the header and data encryption operations, respectively.
\item \texttt{ENCHDR}: The complete original FITS header, encrypted with the Header Key.
\item \texttt{AUTHTAG\_H}: The 128-bit GCM authentication tag verifying the integrity of \texttt{ENCHDR}.
\item \texttt{ORIG\_*}: Plaintext copies of essential structural keywords (e.g., \texttt{ORIG\_NAXIS}, \texttt{ORIG\_BITPIX}) required to parse the file structure before decryption.\footnote{The four \texttt{ORIG\_*} keywords (\texttt{ORIG\_BITPIX}, \texttt{ORIG\_NAXIS}, \texttt{ORIG\_NAXIS1}, \texttt{ORIG\_NAXIS2}) are stored in the encrypted container under the abbreviated forms \texttt{ORIG\_BPX}, \texttt{ORIG\_NAX}, \texttt{ORIG\_NA1}, \texttt{ORIG\_NA2} to comply with the 8-character keyword name limit of the FITS standard. This abbreviation is internal to the encrypted container; standard keyword names are always written to the decrypted output.}
\item \texttt{AUTHTAG\_D}: The 128-bit GCM authentication tag protecting the encrypted pixel data and its associated metadata context.
\end{itemize}
\item BINTABLE Extension (Data Container): The original pixel array is encrypted with the Data Key and stored as a Binary Large Object (BLOB) in a single-row Binary Table extension.
\end{enumerate}

The encrypted pixel payload is stored as a single-row, single-column binary blob (\texttt{TFORM='<N>B'}) in a BinTableHDU extension; this is the standard FITS mechanism for embedding arbitrary binary content, and is semantically correct because the bytes are not, by construction, an image. Both the encrypted container and all decryption outputs are opened by Astropy and cfitsio without warnings or errors, recognizing the structure as syntactically valid FITS files.

The encrypted container additionally writes the FITS-standard \texttt{CHECKSUM} and \texttt{DATASUM} keywords to both HDUs, computed via cfitsio's \texttt{fits\_write\_chksum} \citep{seaman_fits_1995, pence_lossless_2009}. These provide accidental-corruption detection (32-bit checksum, broad ecosystem compatibility) complementary to the AES-GCM authentication tags (\texttt{AUTHTAG\_H} and \texttt{AUTHTAG\_D}), which detect intentional manipulation. Both layers are present by design.

\noindent Nonce Generation and Uniqueness Assurance:
The security of the AES-GCM mode relies strictly on the uniqueness of the initialization vector for a given key. To eliminate the risk of nonce reuse, the framework employs a stochastic generation approach compliant with NIST SP 800-38D guidelines \citep{dworkin_recommendation_2007}. The 96-bit nonces are generated utilizing the host system's entropy pool (e.g., via \texttt{/dev/urandom}). Given the 96-bit namespace, the probability of a collision is bounded by the birthday formula $P \approx N^2 / (2 \cdot 2^{96})$, where $N$ is the number of nonces generated under a single key and the 96-bit nonce length is specified by NIST SP 800-38D for AES-GCM. For LSST-scale workloads ($N \sim 10^7$ images per year, single key), this gives $P \approx 6.3 \times 10^{-16}$ per year, negligible by any operational standard. For larger archives reaching $N \sim 10^9$ files per key, $P \approx 6.3 \times 10^{-12}$, providing an acceptable security margin for operational use. Archives approaching $N \sim 2^{32}$ files per key fall into the regime where NIST SP 800-38D \S{}8.3 recommends key rotation; our framework supports this through the explicit key reference required by every encryption operation.

This dual-key design allows a Policy Engine to grant access to the header (e.g., for query validation) without releasing the key for the scientific pixel data.

\subsubsection{Context Binding via Associated Data (AAD)}
A unique feature of the implementation is the use of GCM's Associated Data (AAD) capability to cryptographically bind the plaintext metadata to the encrypted pixels. The library constructs the AAD vector for the data payload by concatenating the data Nonce with the string representation of the critical structural keywords:

\begin{multline*}
\text{AAD}_{Data} = \text{NONCE}_D \parallel \text{Serialize}(\text{ORIG\_BITPIX}) \\
\parallel \text{Serialize}(\text{ORIG\_NAXIS}) \parallel \dots
\end{multline*}

\noindent where the $\text{Serialize}(\cdot)$ function converts numerical values to their strict ASCII decimal representation (e.g., the integer $-32$ becomes the byte sequence \texttt{'-32'}), eliminating potential ambiguities arising from padding spaces or formatting variances.

This AAD is fed into the GHASH computation during both encryption and decryption. While the \texttt{ORIG\_*} keywords remain readable in the header to allow standard FITS tools to identify the file size, any tampering with these values (e.g., altering the coordinate system or exposure time) causes the computed hash to diverge from \texttt{AUTHTAG\_D}. This prevents ``context swapping attacks,'' ensuring the scientific data cannot be presented with falsified telemetry.

\paragraph{Application Programming Interface}
The library exposes a C-style API designed for integration with high-level languages like Python. As shown in Table~\ref{tab:api}, the interface abstracts the complexity of GPU memory management and GCM tag verification.

\begin{table}[t]
\centering
\caption{Core functions and data structures of the \texttt{GpuFitsCrypt} library. The API is designed to return granular status codes, enabling applications to distinguish between full access, partial access (metadata only), and security failures.}
\label{tab:api}
\begin{tabularx}{\columnwidth}{l X}
\toprule
\textbf{Component} & \textbf{Description} \\
\midrule
\multicolumn{2}{c}{\textit{Functions}} \\
\texttt{gfc\_context\_create} & Initializes session: pre-allocates pinned memory and CUDA streams. \\
\texttt{gfc\_decrypt\_frame}  & Decrypts and authenticates a FITS frame. Returns \texttt{FitsOperationResult}. \\
\texttt{gfc\_context\_destroy} & Teardown: releases all GPU and host resources. \\
\midrule
\multicolumn{2}{c}{\textit{Data Structures: \texttt{FitsOperationResult}}} \\
\texttt{data\_buffer} & Pointer to the decrypted FITS file in memory (ready for Astropy). \\
\texttt{error\_code} & 0: Success (check warning); $<$0: System/IO Error. \\
\texttt{warning\_code} & \textbf{0}: Full Access; \textbf{1}: Header Fallback; \textbf{2}: Payload Suppressed (Data Key mismatch). \\
\bottomrule
\end{tabularx}
\end{table}

\subsection{GPU Acceleration Strategy}

Achieving high throughput with AES-GCM on large files requires masking the latency of memory transfers and overcoming the inherently serial nature of the authentication phase.

\subsubsection{Parallel GHASH Reduction}
\label{subsubsec:ghash_theory}
The standard GHASH algorithm operates via a serial chain defined by the recurrence relation:

\begin{equation}
    Y_i = (Y_{i-1} \oplus X_i) \cdot H
    \label{eq:ghash_serial}
\end{equation}

\noindent where $Y_i$ is the 128-bit authentication tag accumulator at step $i$, $Y_{i-1}$ is the previous state, $X_i$ represents the current 128-bit input block (which can be either Associated Data or Ciphertext), $\oplus$ denotes the bitwise XOR operation, and $\cdot$ represents multiplication in the Galois Field $GF(2^{128})$ using the precomputed hash subkey $H$.

\begin{figure*}[htbp]
    \centering
    \begin{tikzpicture}[
    node distance=0.8cm and 0.8cm,
    >=latex, 
    thick,
    block/.style={
        rectangle, 
        draw, 
        fill=blue!10, 
        minimum size=0.8cm, 
        minimum height=0.8cm,
        font=\small,
        outer sep=0pt
    },
    op/.style={
        circle, 
        draw, 
        inner sep=0pt, 
        minimum size=0.5cm, 
        fill=white, 
        font=\small
    },
    const/.style={
        circle, 
        draw, 
        inner sep=1pt, 
        fill=gray!10, 
        font=\footnotesize, 
        minimum size=0.6cm
    }
]

\def\titley{6.2}

\node (s_label) at (2.0, \titley) {\textbf{(a) Serial GHASH ($O(N)$)}};

\node[block] (x1) at (0,0) {$X_1$};
\node[block] (x2) [right=0.8cm of x1] {$X_2$};
\node[block] (x3) [right=0.8cm of x2] {$X_3$};
\node[block] (x4) [right=0.8cm of x3] {$X_4$};

\node (y_in) [left=0.6cm of x1, yshift=0.8cm] {$Y_{i-1}$};
\node[op] (xor1) [above=0.45cm of x1] {$\oplus$};
\node[op] (mul1) [above=0.45cm of xor1] {$\otimes$};
\node[const] (h1) [left=0.3cm of mul1] {$H$};

\draw[->] (x1) -- (xor1);
\draw[->] (y_in) -- (xor1);
\draw[->] (xor1) -- (mul1);
\draw[->] (h1) -- (mul1);

\node[op] (xor2) at (x2 |- mul1) {$\oplus$}; 
\node[op] (mul2) [above=0.45cm of xor2] {$\otimes$};
\node[const] (h2) [left=0.3cm of mul2] {$H$};

\draw[->] (mul1) -- (xor2);
\draw[->] (x2) -- (xor2);
\draw[->] (xor2) -- (mul2);
\draw[->] (h2) -- (mul2);

\node[op] (xor3) at (x3 |- mul2) {$\oplus$};
\node[op] (mul3) [above=0.45cm of xor3] {$\otimes$};
\node[const] (h3) [left=0.3cm of mul3] {$H$};

\draw[->] (mul2) -- (xor3);
\draw[->] (x3) -- (xor3);
\draw[->] (xor3) -- (mul3);
\draw[->] (h3) -- (mul3);

\node[op] (xor4) at (x4 |- mul3) {$\oplus$};
\node[op] (mul4) [above=0.45cm of xor4] {$\otimes$};
\node[const] (h4) [left=0.3cm of mul4] {$H$};

\draw[->] (mul3) -- (xor4);
\draw[->] (x4) -- (xor4);
\draw[->] (xor4) -- (mul4);
\draw[->] (h4) -- (mul4);

\draw[->] (mul4) -- +(0,0.5) node[above] {$Y_{final}$};

\begin{scope}[xshift=8.5cm]

\node (p_label) at (2.5, \titley) {\textbf{(b) Parallel Tree GHASH ($O(\log N)$)}};

\node[block] (px1) at (0,0) {$X_1$};
\node[block] (px2) [right=0.5cm of px1] {$X_2$};

\node[block] (px3) [right=1.6cm of px2] {$X_3$}; 
\node[block] (px4) [right=0.5cm of px3] {$X_4$};

\node[op] (pmul1) [above=0.8cm of px1] {$\otimes$};
\node[const] (ph1) [left=0.3cm of pmul1] {$H^{2^0}$};
\node[op] (pxor1) at (px2 |- pmul1) {$\oplus$};

\draw[->] (px1) -- (pmul1);
\draw[->] (ph1) -- (pmul1);
\draw[->] (pmul1) -- (pxor1);
\draw[->] (px2) -- (pxor1);

\node[op] (pmul2) [above=0.8cm of px3] {$\otimes$};
\node[const] (ph2) [left=0.3cm of pmul2] {$H^{2^0}$};
\node[op] (pxor2) at (px4 |- pmul2) {$\oplus$};

\draw[->] (px3) -- (pmul2);
\draw[->] (ph2) -- (pmul2);
\draw[->] (pmul2) -- (pxor2);
\draw[->] (px4) -- (pxor2);

\node[op] (pmul3) [above=1.2cm of pxor1] {$\otimes$};
\node[const] (ph3) [left=0.3cm of pmul3] {$H^{2^1}$};
\node[op] (pxor3) at (pxor2 |- pmul3) {$\oplus$};

\draw[->] (pxor1) -- (pmul3);
\draw[->] (ph3) -- (pmul3);
\draw[->] (pmul3) -- (pxor3);
\draw[->] (pxor2) -- (pxor3);

\node[op] (pmul_final) [above=0.8cm of pxor3] {$\otimes$};
\node[const] (ph_final) [left=0.3cm of pmul_final] {$H$};

\draw[->] (pxor3) -- (pmul_final);
\draw[->] (ph_final) -- (pmul_final);
\draw[->] (pmul_final) -- +(0,0.6) node[above] {$Y_{final}$};

\end{scope}

\end{tikzpicture}
    
    \caption{Comparison of GHASH strategies. \textbf{(a)} The standard serial approach creates a dependency chain where each step waits for the previous one. \textbf{(b)} The proposed parallel approach uses a binary reduction tree. By precomputing powers of $H$ ($H^{2^0}, H^{2^1}, \dots$), independent blocks can be combined in parallel steps ($O(\log N)$), saturating the GPU cores.}
    \label{fig:ghash_comparison}
\end{figure*}
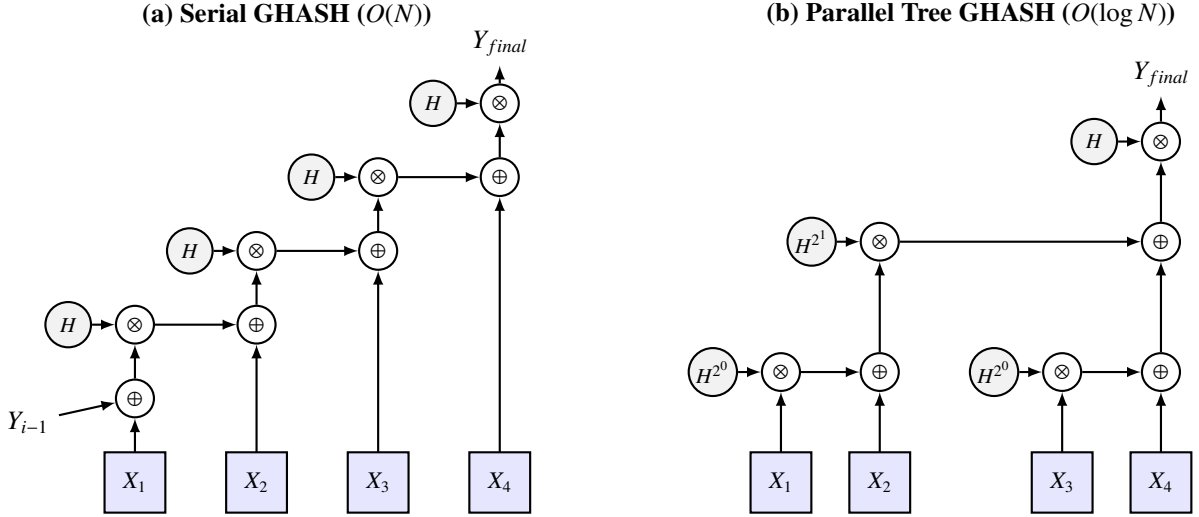

This serial dependency implies that the processing of block $i$ cannot strictly begin until block $i-1$ is complete (see Figure~\ref{fig:ghash_comparison}-a), making it typically ill-suited for the massive parallelism of GPUs. To address this, the implementation utilizes a parallel reduction strategy operating in the Galois Field $GF(2^{128})$, inspired by recent optimizations for parallel architectures \citep{lee_parallel_2025}.

The library precomputes a table of powers of the hash key $H$ ($H^{2^0}, H^{2^1}, H^{2^2}, \dots, H^{2^k}$). This sequence corresponds to the stride length at each level of a binary reduction tree. As illustrated in Figure~\ref{fig:ghash_comparison}-b, this enables the input data to be split into independent blocks and combined in a hierarchical manner. Instead of a linear dependency chain, GPU threads compute partial hashes concurrently, multiplying by the appropriate power of $H$ to "jump" gaps in the sequence. This reduces the computational complexity from linear time $O(N)$ to logarithmic time $O(\log N)$, effectively matching the throughput of the counter-mode encryption component.

\subsubsection{Asynchronous Streams Pipeline}
The library implements an asynchronous pipeline using NVIDIA CUDA streams to overlap host-device communication with computation:

\begin{enumerate}
\item Dual-Stream Context: The library maintains a persistent context with two concurrent CUDA streams: \texttt{streamH} (Header) and \texttt{streamD} (Data), along with Pinned Memory (page-locked) host buffers to maximize PCIe bandwidth.
\item Header Processing: \texttt{streamH} handles the \texttt{ENCHDR} decryption. Since headers are typically small ($<80$ kB), this provides low-latency access to metadata.
\item Data Pipeline: \texttt{streamD} manages the heavy lifting for the pixel data. The host reads the encrypted BLOB directly into pinned memory. The stream then executes the GCM pipeline:
\begin{itemize}
\item H-Power Generation: Computes powers of the authentication key on the fly if the data size requires it.
\item AES-CTR Kernel: Generates the keystream and decrypts the payload in parallel blocks.
\item GHASH Kernel: Performs the parallel reduction on the ciphertext and AAD to verify integrity.
\end{itemize}
\item Validation Gate: The library enforces a strict validation gate. The output buffer is only marked as valid if the locally computed GHASH matches the \texttt{AUTHTAG} stored in the file. If verification fails, the operation returns a distinct integrity error code, ensuring no corrupted or manipulated data is released to the user.
\end{enumerate}

\section{Secure and Transparent Data Access}
\label{sec:secure_access}

The primary goal of the framework is to provide robust security and granular access control without imposing a significant burden on the end-user. This is achieved through a combination of a transparency mechanism at the application layer and a strict policy enforcement workflow. This section describes the user interaction model and how the dual-key FITS format enables diverse and efficient access patterns.

\subsection{User Interaction and Transparency Mechanism}
\label{subsec:user_interaction}

From the perspective of a research scientist, interacting with the encrypted data archive must be seamless. The complexity of decryption, key handling, and policy checks is abstracted away by the framework's interface. The implementation provides a C-style API (refer to Table~\ref{tab:api}) designed to be wrapped by higher-level languages commonly used in astronomy, such as Python.

Listing~\ref{lst:python_example} (~\ref{app:code_listings}) demonstrates a typical user workflow. The user initializes a session context once at the beginning, pre-allocating necessary GPU and pinned memory resources via \texttt{gfc\_context\_create}. For each encrypted FITS file, the \texttt{gfc\_decrypt\_frame} function is called. Upon success, the library returns a \texttt{FitsOperationResult} structure containing a pointer to a fully decrypted FITS file in memory. This memory buffer can be passed directly to standard scientific libraries like \texttt{Astropy} using memory-efficient methods such as \texttt{HDUList.fromstring()}, treating the decrypted data as a virtual file. This approach achieves a high degree of transparency; the analysis code interacts with standard data objects, remaining agnostic to the underlying cryptographic operations.

\subsection{Policy Enforcement and Granular Access}
\label{subsec:policy_enforcement}

The framework's ability to enforce fine-grained access control is predicated on the dual-key design of the encrypted FITS format, orchestrated by the architecture's Control Plane. As depicted in the workflow diagram (Figure~\ref{fig:policy_workflow_diagram} in Section~\ref{sec:architecture}), abstract policy rules are translated into concrete cryptographic operations through a conditional key release mechanism.

Consider a scenario involving a dataset under a proprietary period. Table~\ref{tab:policy_examples} illustrates how different user roles and policies translate into the specific keys provisioned by the Access Control Module (ACM) and the resulting data access level provided by the library.

\begin{table*}[t]
\centering
\caption{Role-based and temporal access control policies. The table reflects the full granularity of the system, demonstrating how different key combinations trigger specific behaviors (Full Access, Suppression, or Fallback).}
\label{tab:policy_examples}
\begin{tabularx}{\textwidth}{l X l X}
\toprule
\textbf{Scenario / Role} & \textbf{Policy Context} & \textbf{KMS Key Provisioning} & \textbf{Resulting Access / System Behavior} \\
\midrule
\textbf{Principal Inv.} & \texttt{FULL\_ACCESS} & Header + Data Key & \textbf{Full Access}: Metadata and Pixels fully decrypted and authenticated. \\
\midrule
\textbf{Archive Miner} & \texttt{METADATA\_ONLY} & Header Key Only & \textbf{Payload Suppression}: Header authenticated; Pixel data physically zeroed to protect IP. \\
\midrule
\textbf{Blind Analysis} & \texttt{PIXELS\_ONLY} (Sensitive Telemetry) & Data Key Only & \textbf{Header Fallback}: Minimal header reconstructed from plaintext; Pixels decrypted. \\
\midrule
\textbf{Integrity Check} & \texttt{role:any} on \texttt{data:tampered} & Header + Data Key & \textbf{Auth Failure}: GHASH mismatch triggers suppression. Safety mechanism active. \\
\bottomrule
\end{tabularx}
\end{table*}

A critical feature of the Policy Engine is its support for time-based rules, essential for automating data embargoes. As shown in Table~\ref{tab:policy_examples}, a ``Public User'' is initially denied access. However, a rule such as \texttt{allow role:public if data.release\_date < now()} automatically transitions the access level upon the embargo's expiration. After this date, the same request results in the provisioning of both keys, granting \texttt{FULL\_ACCESS}.

Furthermore, the model allows decoupling metadata access from pixel data access. A collaborator may be granted the Header Key to decrypt the \texttt{ENCHDR} keyword, enabling analysis of observational telemetry without accessing the scientific pixel data. Conversely, to support blind analysis scenarios where metadata is restricted, the library implements an automatic fallback mechanism. If the Header Key is denied (or incorrect), the library constructs a structurally valid, minimal FITS header in memory using the unencrypted \texttt{ORIG\_*} keywords (e.g., \texttt{ORIG\_NAXIS}, \texttt{ORIG\_BITPIX}). This ensures that even with restricted permissions, the user receives a parsable FITS file, preserving compatibility with standard tools while keeping the core scientific content secure.

Crucially, unlike confidentiality-only modes such as AES-CTR, the use of AES-GCM ensures data integrity through compartmentalized verification. In our implementation, the integrity checks for metadata and pixel data are decoupled. Consequently, a mismatch in the Data Key (or integrity failure) triggers a payload suppression mechanism: the library returns a structurally valid FITS object where the scientific pixel array is zero-filled (physically suppressed via \texttt{memset}), while the authenticated header remains fully accessible. 

This mechanism, combined with the automatic header fallback described above, ensures that granular access policies are enforced without crashing the read operation or exposing the application to corrupted content, maintaining compatibility with standard FITS viewers even in restricted access scenarios. Empirical validation of all four access scenarios on a synthetic $256\times256$ float32 FITS file with representative WCS metadata confirms that each scenario produces a syntactically valid FITS output (full content, zero-filled payload, or minimal header), opens cleanly in Astropy without warnings, and is bit-exact to the original in the Full Access case.

\section{GPU-Accelerated Cryptography Implementation and Performance}
\label{sec:implementation_performance}

This section details the technical implementation of the \texttt{GpuFitsCrypt} module and presents a comprehensive performance analysis. The discussion focuses on the specific strategies employed to parallelize the authenticated encryption process on the GPU and the resulting throughput across consumer and datacenter hardware tiers.

\subsection{Choice of Cryptographic Mode}
\label{subsec:crypto_mode}

We selected AES-128-GCM (Galois/Counter Mode) as the cryptographic standard. AES-128 is preferred over AES-256 for the threat model of scientific archives with embargo periods of months to a few years: the best known cryptanalysis, the biclique attack of \citet{hutchison_biclique_2011}, requires approximately $2^{126}$ operations, which remains computationally infeasible by orders of magnitude. This is also the choice made by \citet{lee_speed_2024} and \citet{lee_parallel_2025}, whose implementation we adopt as our cryptographic core. The architectural choice does not preclude future migration to a wider block cipher: GpuFitsCrypt isolates the AES core behind a clear interface, and recent results \citep{malal_first_2026} indicate that migration to WAES-256 / Rijndael-256 produces no observable slowdown (3,053.5~Gbps for WAES-256 versus 3,062.5~Gbps for AES-256 on RTX~4090, a difference of approximately 0.3\%). We therefore consider the migration path straightforward when NIST completes the standardization of Rijndael-256 GCM/GMAC currently under public comment.

While Counter (CTR) mode offers high parallelizability and confidentiality, it lacks intrinsic integrity verification, leaving scientific data vulnerable to ciphertext manipulation (e.g., bit-flipping attacks) unless paired with a separate Message Authentication Code (MAC). An alternative encrypt-then-MAC construction would combine AES-CTR with HMAC-SHA256; we chose AES-GCM over this option for three reasons specific to the GpuFitsCrypt deployment context. First, single-pass throughput: CTR+HMAC requires two sequential passes over the data (encryption then MAC), which doubles memory traffic on memory-bandwidth-bound GPU architectures, whereas AES-GCM performs encryption and authentication in a single fused pass. Second, native AAD binding: GpuFitsCrypt cryptographically binds the FITS metadata (size, geometry, exposure parameters) to the encrypted pixel payload through GCM's Associated Authenticated Data field, preventing context-swapping attacks; CTR+HMAC has no native AAD construct, and replicating this guarantee with a custom HMAC scope would require an ad-hoc, non-standardized construction. Third, standardization: AES-GCM is specified by NIST SP~800-38D \citep{dworkin_recommendation_2007} as an approved AEAD mode, whereas CTR+HMAC, although built from approved primitives, is not itself a standardized AEAD construction. CTR+HMAC remains a valid alternative for genuinely streaming workflows where incremental tag verification is the dominant requirement, but for the whole-file retrieval pattern that dominates GpuFitsCrypt's intended deployment, the GCM advantages are decisive. The primary challenge addressed in this work is the efficient parallelization of GCM's authentication component, GHASH, which is inherently sequential.

\subsection{GPU Implementation: Parallel GHASH Architecture}
\label{subsec:gpu_implementation}

The \texttt{GpuFitsCrypt} library utilizes the NVIDIA CUDA platform. Building upon the high-performance bit-sliced AES-CTR core architecture described by \citet{lee_speed_2024}, we introduced a critical modification to ensure cryptographic interoperability across heterogeneous hardware. The original algorithm's counter generation logic was coupled to the GPU grid geometry (total thread count), causing the keystream to vary with hardware configuration. We refactored the kernel to enforce a deterministic, geometry-agnostic mapping between data blocks and AES counters. This ensures that a FITS file encrypted on an edge device (e.g., Jetson Nano) can be correctly decrypted on a datacenter GPU (e.g., NVIDIA A100), regardless of the parallelism degree, without compromising the high-throughput performance. One of the core innovations of this work lies in transforming the serial GHASH dependency chain (Equation~\ref{eq:ghash_serial}) into a parallelizable workload suitable for massive datasets.

\subsubsection{Parallel Tree-Reduction Strategy}
We implement the reduction strategy operating in the Galois Field $GF(2^{128})$ as theoretically defined in Section~\ref{sec:architecture}.

\begin{itemize}
    \item Precomputation: The library generates the look-up table of hash key powers (corresponding to the stride lengths $H^{2^k}$ described in Section~\ref{subsubsec:ghash_theory}) on the GPU. This pre-calculated table allows any data block in the sequence to be processed independently, effectively bridging the positional gap without serial dependencies.
    \item Tree Reduction Kernel: Our custom kernel, \texttt{ghash\_parallel\_reduction}, leverages Shared Memory to load data chunks, minimizing global memory latency. It then performs the synchronized tree reduction (visualized in Figure~\ref{fig:ghash_comparison}-b). In each step, threads combine partial results using the precomputed powers, reducing the complexity from linear $O(N)$ to logarithmic $O(\log N)$.
\end{itemize}

\subsubsection{Scalable MapReduce for Massive Files}
Since a single CUDA block cannot process an arbitrary number of data blocks due to shared memory limits, we employ a two-phase MapReduce architecture to handle large FITS files (e.g., 600~MB to 3~GB):
\begin{enumerate}
\item Map Phase (Partial Reduction): A grid of CUDA blocks is launched. Each block processes a fixed segment of the input data (ciphertext + AAD) independently, producing a single 16-byte partial hash.
\item Reduce Phase (Final Combination): A secondary kernel, \texttt{combine\_partial\_ghash}, takes the array of partial hashes and combines them on the GPU. This architecture prevents the latency of transferring intermediate results back to the CPU, keeping the entire authentication pipeline on the device.
\end{enumerate}

\subsubsection{Optimized Memory Pipeline}
To ensure the GPU is fed data at a rate matching its compute capability, the implementation utilizes:
\begin{itemize}
\item Pinned Memory I/O: All host buffers use page-locked memory (\texttt{cudaMallocHost}) to maximize PCIe bandwidth and enable Direct Memory Access.
\item Dual-Stream Overlap: As detailed in Section \ref{sec:architecture}, encryption/decryption (AES-CTR) and authentication (GHASH) are managed within a stream pipeline that overlaps with disk I/O operations.
\end{itemize}

\subsection{Performance Benchmarking}
\label{subsec:benchmarking}

We evaluated the framework on a diverse set of hardware, ranging from edge/mobile GPUs to datacenter accelerators (see  Table~\ref{tab:hardware_specs_detailed} in ~\ref{app:benchmarks} for hardware specs). The dataset consisted of both real and synthetic FITS images ranging from 16~MB to approximately 600~MB (Large) and 3.4~GB (Extra-Large) (see Table~\ref{tab:dataset_specs}).

\subsubsection{Experimental Protocol}
To ensure statistical robustness and reproducibility, a rigorous measurement protocol was applied to all reported benchmarks:

\begin{itemize}
    \item Execution Context: The CUDA context, including pinned memory allocation and stream creation, was initialized once per session and reused across all runs. This ensures that measurements reflect the steady-state performance of the pipeline, excluding the one-time overhead of driver initialization and context creation.
    \item Sampling Strategy: For every hardware and file size configuration, we performed 5 warm-up runs (discarded) to stabilize GPU clock states and JIT compilation effects, followed by 200 timed repetitions. Reported values represent the median of these executions, accompanied by standard deviations to quantify variability.
    \item Timing Mechanisms: GPU kernel execution times were measured using high-precision \texttt{cudaEventRecord} pairs on the device stream. To isolate raw computational throughput accurately, these markers were placed strictly around the cryptographic kernels, explicitly excluding host-device memory transfers and auxiliary operations. Host-side End-to-End latencies were recorded using monotonic system timers (e.g., \texttt{time.monotonic()} in Python) to capture the full wall-clock duration experienced by the user.
\end{itemize}

\subsubsection{Performance Metrics Definition}
To accurately characterize the system's behavior across different bottlenecks, we define three distinct metrics used in the analysis:

\begin{enumerate}
    \item Kernel Time: Represents the exclusive execution duration of the decryption and authentication kernels on the GPU. This metric isolates raw computational throughput from system I/O.
    \item End-to-End (E2E) Time: Measures the total time from the initial API call until the decrypted data is fully available in system RAM. This includes disk I/O, host-to-device PCIe transfers, kernel execution, and Python runtime overhead.
    \item Net Cryptographic Cost: Defined as $(\text{E2E}_{\text{GCM}} - \text{Astropy}_{\text{Load}})$. This metric isolates the computational overhead of the security layer by subtracting the baseline I/O cost of loading a standard unencrypted FITS file via \texttt{Astropy}. It serves as a proxy for the "price of security" in a production pipeline.
\end{enumerate}

\subsection{Results and Analysis}
\label{subsec:results}

\paragraph{Kernel Parameter Optimization and Scalability Regimes}
Performance is sensitive to the Bit-Sliced AES parameters: \emph{Thread Size} (TS) and \emph{Repeat Blocks} (R). An exhaustive search was performed for each hardware configuration.
Table~\ref{tab:optimal_kernel_configs} presents the kernel-only execution times for a standard "Large" astronomical image ($\approx 600$~MB).

\begin{table}[t]
\centering
\caption{Kernel-optimal parameters (\texttt{TS}: Thread Size, \texttt{R}: Repeat Blocks) and execution time for AES-GCM on the large FITS dataset ($\approx 600$~MB). Note that for this workload size, the execution time on Datacenter GPUs (H100, A100) is dominated by kernel launch latency rather than computational throughput, resulting in times comparable to consumer cards.}
\label{tab:optimal_kernel_configs}
\resizebox{\columnwidth}{!}{%
\begin{tabular}{l c c c}
\toprule
\textbf{GPU Model} & \textbf{Opt. TS} & \textbf{Opt. R} & \textbf{Kernel Time (s)} \\
\midrule
\textit{Datacenter / HPC} \\
H100 PCIe & 256 & 8 & 0.026 $\pm$ 0.000 \\
A100 & 256 & 8 & 0.042 $\pm$ 0.001 \\
L40S & 64 & 8 & 0.039 $\pm$ 0.000 \\
\midrule
\textit{Consumer / Workstation} \\
GeForce RTX 3090 & 128 & 8 & 0.070 $\pm$ 0.000 \\
GeForce RTX 3060 & 128 & 4 & 0.236 $\pm$ 0.001 \\
GeForce RTX 3050 Ti & 128 & 64 & 0.369 $\pm$ 0.000 \\
\midrule
\textit{Edge} \\
Jetson Orin & 64 & 64 & 0.542 $\pm$ 0.003 \\
\bottomrule
\end{tabular}%
}
\end{table}

A counter-intuitive observation from Table~\ref{tab:optimal_kernel_configs} is the parity between high-end datacenter GPUs (H100) and consumer hardware (RTX 3060) for the 600~MB dataset. This phenomenon indicates that for single-image processing of this size, the workload operates in a latency-bound regime. The massive parallelism of the H100 is underutilized, and the total time is dominated by constant system overheads (PCIe latency, kernel launch time, and I/O initialization) rather than raw arithmetic throughput.

To mitigate the impact of system overheads and evaluate the architectures in a throughput-bound regime, we extended the benchmark to an "Extra-Large" dataset ($\approx 3.4$~GB). This workload not only stresses the computational units but also tests the memory capacity limits of the devices. Table~\ref{tab:gcm_vs_ctr_comparison} details the End-to-End (E2E) performance for both file sizes.

\begin{table*}[t]
\centering
\caption{Performance scalability analysis comparing Large ($\approx 600 MB$) and Extra-Large ($\approx 3.4 GB$) FITS files. The table decomposes the total End-to-End (E2E) time into a baseline I/O component (\texttt{Astropy Time}) and the \texttt{Net GCM Cost} (calculated as E2E Time - Astropy Time). This isolation reveals that while I/O masks performance differences on smaller files, the \texttt{Net GCM Cost} scales effectively on datacenter hardware for larger workloads. \textit{Note: Consumer-grade and edge devices with limited VRAM (RTX 3060, RTX 3050 Ti, and Jetson Orin) are excluded from the Extra-Large benchmark as the dataset size exceeds their available memory capacity.}}
\label{tab:gcm_vs_ctr_comparison}
\resizebox{\textwidth}{!}{%
\begin{tabular}{l c c c c c c}
\toprule
\textbf{GPU Model} & \textbf{Astropy Time} & \textbf{E2E CTR Time} & \textbf{E2E GCM Time} & \textbf{Net CTR Cost} & \textbf{Net GCM Cost} & \textbf{GCM/CTR Net Ratio} \\
\midrule
\textit{Large ($\approx 600 MB$)} \\
H100 PCIe          & 0.122 & 0.482 & 1.344 & 0.359 & 1.222 & 3.40x \\
A100               & 0.242 & 0.523 & 1.293 & 0.281 & 1.051 & 3.74x \\
L40S               & 0.200 & 0.563 & 1.406 & 0.363 & 1.206 & 3.32x \\
GeForce RTX 3090   & 0.178 & 0.542 & 1.316 & 0.364 & 1.138 & 3.13x \\
GeForce RTX 3060   & 0.101 & 0.711 & 1.296 & 0.611 & 1.196 & 1.96x \\
GeForce RTX 3050 Ti & 0.123 & 0.932 & 1.513 & 0.809 & 1.390 & 1.72x \\
Jetson Orin        & 0.225 & 0.698 & 1.738 & 0.473 & 1.513 & 3.20x \\
\midrule
\textit{Extra-Large ($\approx 3.4 GB$)} \\
H100 PCIe          & 1.460 & 2.798 & 8.000 & 1.338 & 6.541 & 4.89x \\
A100               & 1.393 & 3.005 & 7.601 & 1.612 & 6.208 & 3.85x \\
L40S               & 1.715 & 3.310 & 7.945 & 1.595 & 6.230 & 3.91x \\
GeForce RTX 3090   & 1.048 & 3.086 & 7.215 & 2.038 & 6.167 & 3.03x \\
\bottomrule
\end{tabular}%
}
\end{table*}

\paragraph{The Cost of Integrity: GCM vs. CTR}
Comparing the \texttt{Net GCM Cost} against \texttt{Net CTR Cost} allows us to isolate the computational overhead of the authentication layer. Across all architectures, GCM introduces a slowdown factor ranging from $1.72$x to $4.89$x compared to unauthenticated encryption. This reflects the unavoidable arithmetic cost of the finite field multiplications required for GHASH. The H100 PCIe ratio increases from $3.40$x in the Large regime to $4.89$x in the Extra-Large regime, reflecting the transition from an I/O-dominated to a compute-dominated regime as the dataset size approaches the cryptographic kernel's saturation point. This transition is the expected behavior for an architecture optimized for high-throughput compute, and it is consistent across the four GPUs benchmarked in the Extra-Large regime (range: $3.03$x to $4.89$x).

\paragraph{Comparison with CPU Baselines}
While we focus on characterizing the GPU performance envelope, placing these results in context with CPU-based execution is essential. Recent state-of-the-art benchmarks for fully parallelized AES-GCM on GPUs \citep{lee_parallel_2025} demonstrate a speedup factor of $\approx 15\times$ compared to highly optimized multi-threaded CPU implementations (Intel Core i7 with AES-NI instructions using OpenSSL). Furthermore, the same study highlights a power efficiency advantage of over $20\times$ for the GPU architecture. Our results are consistent with this performance gap, validating the architectural choice of accelerator-based cryptography for massive datasets. Within the GPU tier, prior optimized AES-CTR implementations have achieved approximately 878~Gbps on consumer hardware \citep{tezcan_optimization_2021}, providing additional context for the GPU cryptographic performance envelope demonstrated here.

\paragraph{Comparison with GPU-Accelerated Literature}
Table~\ref{tab:literature_comparison} places GpuFitsCrypt in the context of existing GPU-accelerated AES implementations. The most direct comparison is with \citet{lee_parallel_2025}, who measure GHASH-only throughput using 16\,KB pre-staged message blocks, a scope equivalent to our \texttt{GHashKernelOnly} measurement. On the RTX 3090, GpuFitsCrypt achieves 472\,Gbps versus 247\,Gbps reported by \citeauthor{lee_parallel_2025}, a factor of approximately 1.91$\times$ higher. On Ada Lovelace hardware, the L40S reaches 664\,Gbps, exceeding the 536\,Gbps reported for the RTX 4090 by approximately 24\%. These differences reflect our bit-sliced implementation, which operates on $\approx$600\,MB real FITS payloads rather than synthetic 16\,KB blocks, and may also reflect architectural and clock-speed differences between GPU models. While these raw throughput metrics provide an essential baseline, GpuFitsCrypt extends beyond standalone cryptographic kernels by integrating full AES-GCM authenticated encryption with Associated Authenticated Data binding and direct integration with the standard astronomical Python ecosystem (Astropy/cfitsio), producing FITS-compatible encrypted containers that open cleanly in unmodified scientific tools.

\begin{table*}[t]
\centering
\caption{Comparison of GpuFitsCrypt kernel-level throughput against prior GPU AES implementations from the literature. Throughput (Gbps) is kernel-only, excluding host-side I/O. GpuFitsCrypt values are GHASH-only medians over $N=200$ runs on the Large ($\approx$600\,MB) FITS dataset (TS256\_R4 configuration). Lee et al.\ \citeyearpar{lee_parallel_2025} values are for 16\,KB pre-staged messages; the measurement scope is equivalent (GHASH kernel only) but the input size differs. AAD: support for Associated Authenticated Data; FITS-aware: direct integration with the Astropy/cfitsio ecosystem.}
\label{tab:literature_comparison}
\resizebox{\textwidth}{!}{%
\begin{tabular}{l c c r c c}
\toprule
\textbf{Implementation} & \textbf{GPU} & \textbf{AES Mode} & \textbf{Throughput (Gbps)} & \textbf{AAD} & \textbf{FITS-aware} \\
\midrule
\textit{This work (GpuFitsCrypt)} \\
GpuFitsCrypt & L40S & GCM (GHASH only) & 664 & $\checkmark$ & $\checkmark$ \\
GpuFitsCrypt & H100 PCIe & GCM (GHASH only) & 541 & $\checkmark$ & $\checkmark$ \\
GpuFitsCrypt & RTX 3090 & GCM (GHASH only) & 472 & $\checkmark$ & $\checkmark$ \\
GpuFitsCrypt & A100-SXM4 & GCM (GHASH only) & 436 & $\checkmark$ & $\checkmark$ \\
\midrule
\textit{Prior GPU AES literature} \\
\citet{lee_parallel_2025} & RTX 4090 & GCM (GHASH, 16\,KB) & 536 & $\checkmark$ & $\times$ \\
\citet{lee_parallel_2025} & RTX 3090 & GCM (GHASH, 16\,KB) & 247 & $\checkmark$ & $\times$ \\
\citet{lee_speed_2024} & RTX 3080 & CTR & 1623 & $\times$ & $\times$ \\
\citet{tezcan_optimization_2021} & RTX 2070 Super & CTR & 879 & $\times$ & $\times$ \\
\bottomrule
\end{tabular}%
}
\end{table*}

\begin{figure}[!htbp]
 \centering
 \includegraphics[width=\columnwidth]{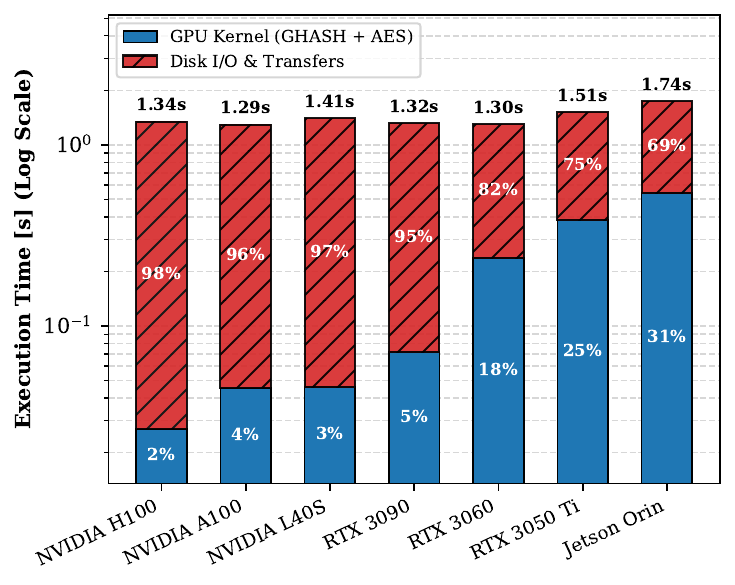}
\caption{
    \textbf{Execution Time Breakdown for AES-GCM (Large File).} 
    The chart illustrates the split between GPU Kernel Computation (Blue) and System/Disk I/O \& Overhead (Red) on a logarithmic scale. 
    For this moderate workload size ($\approx 600$~MB), the process operates in a latency-bound regime, where the GPU kernel accounts for a small fraction of the total execution time.
    This visualizes how system overheads and I/O latency dominate performance when the GPU is not fully saturated, explaining the parity between consumer and datacenter hardware in this specific scenario.
}
 \label{fig:time_breakdown_perc}
\end{figure}

\paragraph{Throughput and Bottleneck Analysis}
Figure~\ref{fig:time_breakdown_perc} illustrates the time breakdown for the AES-GCM operation on the Large dataset. The results highlight a critical architectural insight: for files in the hundreds of megabytes range, the process is latency-bound. The GPU kernel execution time (blue) is a fraction of the total End-to-End time, which is dominated by system overheads and I/O (red).

This observation explains why a consumer-grade RTX 3060 can appear competitive with an H100 for single-image processing: for moderate file sizes, the workflow is dominated by system overheads rather than compute capacity.

In the latency-bound Large regime, the variance in the Astropy I/O baseline dominates the Net GCM Cost ($\textsf{E2E} - \textsf{Astropy}$), which can occasionally mask the raw compute advantage of datacenter architectures like the H100 when compared to high-bandwidth consumer cards like the RTX 3090; Table~\ref{tab:optimal_kernel_configs} confirms the expected hardware ordering in pure-kernel work. In the Extra-Large regime, the expected hardware ordering is confirmed for unauthenticated throughput: the H100 PCIe achieves the lowest Net CTR Cost (1.34\,s) across all benchmarked GPUs, and the Net GCM Cost values cluster within a narrow range (6.2--6.5\,s), confirming that \texttt{GpuFitsCrypt} scales effectively with data volume.

\paragraph{Scientific Validation via Photometry}
To verify that the cryptographic process preserves the scientific utility of the data beyond mere bitwise integrity, we performed a photometric analysis using GPUPHOT, a high-performance photometry package currently under development (Alarcon and Lemes-Perera, in prep.). GPUPHOT integrates a novel set of convolution-based algorithms optimized for CUDA-enabled GPUs to accelerate point source photometry. We utilized a real scientific exposure from the Two-meter Twin Telescope (TTT3) containing a dense stellar field.

\begin{figure*}[htbp]
    \centering
    \includegraphics[width=\textwidth]{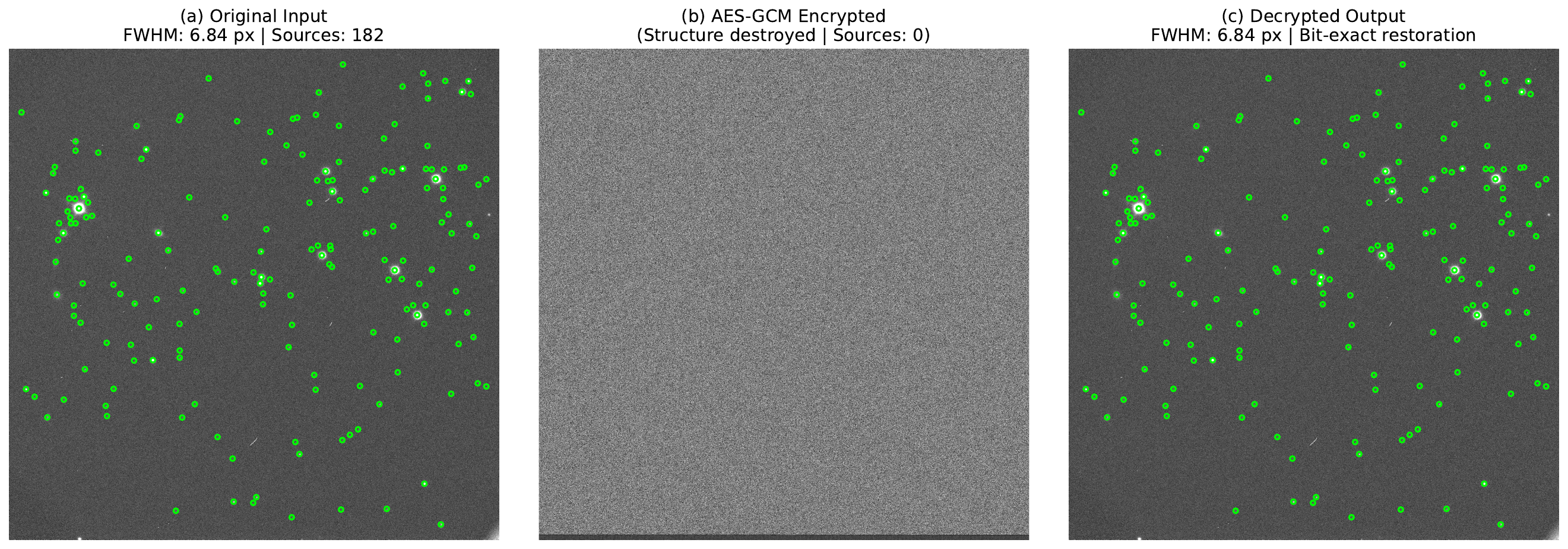}
    \caption{Visual validation of confidentiality and integrity using real scientific data. 
    The test utilizes a raw exposure from the Two-meter Twin Telescope (TTT3). 
    \textbf{(a)} The original FITS image allows the GPUPHOT pipeline to detect 182 point sources with a mean FWHM of 6.84 pixels (green apertures). 
    \textbf{(b)} The AES-GCM encrypted payload appears as high-entropy noise; the photometry software fails to identify any structure (0 sources detected), confirming semantic confidentiality. 
    \textbf{(c)} The decrypted image is bitwise identical to the original, recovering the exact source catalog with identical photometric parameters.}
    \label{fig:scientific_validation}
\end{figure*}

As illustrated in Figure~\ref{fig:scientific_validation}, the validation confirms both security and utility. When applied to the encrypted payload (Panel b), the source extraction algorithm failed to converge, detecting zero sources due to the complete destruction of spatial structure. Conversely, the decrypted output (Panel c) yielded a source catalog identical to the original input. We compared critical parameters for all 182 detected sources, including centroid positions, Full Width at Half Maximum (FWHM $\approx 6.84$ px), and instrumental flux, finding zero residuals. This confirms that the framework introduces no numerical artifacts, preserving the exact radiometric properties required for high-precision astronomy.

\section{Discussion and Future Work}
\label{sec:discussion}

The results presented in this study demonstrate that the historical trade-off between rigorous security and high-performance data access in astronomy is no longer an insurmountable barrier. By leveraging the parallel processing capabilities of modern GPUs, we have shown that it is possible to implement authenticated encryption (AES-GCM) at speeds that match the I/O capabilities of standard storage infrastructure.

\subsection{Latency vs. Throughput in Astronomical Pipelines}
Our benchmarking reveals a distinct dichotomy between single-file latency and aggregate throughput. For interactive analysis of individual images (e.g., a researcher opening a 600~MB file), consumer-grade GPUs like the RTX 3090 or even the RTX 3060 provide performance indistinguishable from datacenter hardware, as the process is bound by I/O latency. This is a positive finding, as it lowers the barrier to entry for secure data access on standard workstations.

However, the value of Datacenter architectures (H100, A100) lies in their capacity for massive concurrency. While a single file does not saturate these cards, their massive VRAM and compute throughput allow for the processing of multiple streams simultaneously, a scenario typical of observatory pipelines reducing thousands of exposures per night. The scalability demonstrated in the Extra-Large benchmarks suggests that these architectures can handle petabyte-scale encryption without becoming a bottleneck.

\subsection{Security Implications of the Architecture}

Our framework's reliance on an external Key Management Service (KMS) is a deliberate architectural choice. It decouples the security policy from the data itself. Unlike password-protected ZIP files or static GPG keys, access to a dataset can be revoked instantly by updating the policy in the control plane, without needing to re-encrypt the petabytes of stored data. Furthermore, the use of GCM's integrity check prevents the "silent corruption" of scientific data, a risk in long-term archival storage that simple encryption methods do not address.

Object-layer encryption complements rather than replaces transport-layer protection. Transport-layer protocols such as SSH and TLS protect data in transit between an archive endpoint and a user, but not after the file has been downloaded, replicated to a backup, or staged on a cold-storage tier. GpuFitsCrypt operates at the object layer: each FITS file is independently encrypted and authenticated, and remains so for the lifetime of the file regardless of where or how it is stored. The two mechanisms therefore protect orthogonal threat surfaces, consistent with the layered security model outlined in Section~\ref{sec:data_challenges}. The standard defense-in-depth posture for sensitive scientific data combines transport-layer encryption for data in motion, object-layer encryption for data at rest, and access policy enforcement at decryption time via the framework's dual-key mechanism.

\paragraph{Layered integrity guarantees}
The CHECKSUM/DATASUM convention written by the framework (see Section~\ref{sec:architecture}) and the AES-GCM authentication tags address orthogonal threats: the former detects accidental corruption (bit rot, transmission errors, partial-write failures) with broad ecosystem compatibility, while the latter detects intentional manipulation with cryptographic strength. To provide robust protection against both threat vectors, \texttt{GpuFitsCrypt} integrates both mechanisms simultaneously, ensuring comprehensive integrity verification.

\subsection{Limitations}
The primary limitation of the current implementation is its dependency on the NVIDIA CUDA platform. While this covers a vast majority of HPC and workstation environments in astronomy, it excludes hardware from other vendors (AMD, Intel GPUs). Additionally, the requirement for pinned memory management, while essential for performance, requires careful integration into host applications to avoid memory exhaustion in highly concurrent multi-user environments.

\subsection{Future Directions}
Future work will focus on four key areas:
\begin{enumerate}
\item Portability: Porting the parallel GHASH reduction kernel to open standards like OpenCL or HIP/ROCm to support a wider range of hardware accelerators.
\item Identity Federation: Integrating the Access Control Module with federated identity providers (e.g., EduGain, ORCID) to facilitate seamless authentication for international collaborations.
\item Attribute-Based Encryption: Exploring advanced cryptographic schemes where policies are mathematically embedded into the ciphertext itself, potentially reducing the reliance on a centralized online Policy Engine for every transaction.
\item Block-size and AEAD upgrades: The modular AES core in GpuFitsCrypt isolates the cipher behind a clear interface, enabling future migration to wider block sizes such as Rijndael-256 / WAES-256 once standardized. Recent benchmarks \citep{malal_first_2026} indicate negligible performance penalty for this transition (approximately 0.3\% on RTX 4090). For deployment scenarios requiring nonce-misuse resistance (e.g., federated archives where multiple operators share a key), AES-GCM-SIV \citep{gueron_aes-gcm-siv_2017, gueron_aes-gcm-siv_2018} provides a drop-in alternative with stronger semantic guarantees.
\end{enumerate}

\section{Conclusions}
\label{sec:conclusions}

We have presented a high-performance framework for securing massive astronomical image catalogs. At its core is \texttt{GpuFitsCrypt}, a library that implements a novel parallel tree-reduction algorithm to accelerate AES-128-GCM on GPUs. This approach successfully overcomes the sequential bottleneck of the GHASH function, achieving end-to-end decryption throughputs that saturate standard storage interfaces.

Our architecture enables fine-grained, policy-based access control (e.g., role-based, temporal embargoes) while guaranteeing both the confidentiality and integrity of scientific data. By binding metadata context to the encrypted payload, we prevent tampering attacks that could compromise scientific validity. This work provides a path forward for data centers to enforce rigorous security during proprietary periods without hindering the pace of discovery, ultimately supporting a trusted and robust transition of data into the public domain.

\section*{CRediT authorship contribution statement}
\label{sec:credit}
Samuel Lemes-Perera: Conceptualization, Methodology, Software, Validation, Formal analysis, Investigation, Data Curation, Writing - Original Draft, Visualization.

Miguel R. Alarcón: Methodology, Formal analysis, Visualization, Validation, Writing - Review \& Editing.

Pino Caballero-Gil: Conceptualization, Supervision, Project administration, Writing - Review \& Editing.

Miquel Serra-Ricart: Supervision, Resources, Validation, Writing - Review \& Editing.

\section*{Declaration of competing interest}
\label{sec:doci}
The authors declare that they have no known competing financial interests or personal relationships that could have appeared to influence the work reported in this paper.

\section*{Software Availability}
\label{sec:software_availability}

The source code for the \texttt{GpuFitsCrypt} library, including the GPU kernels and the Python wrapper, is available under an open-source license at \url{https://github.com/slemesp/GpuFitsCrypt}. A persistent archive of the specific version used for the benchmarks presented in this paper will be deposited in Zenodo upon acceptance.

\section*{Acknowledgements}
\label{sec:acknowledgements}

This work was supported by Light Bridges S.L. under an Industrial PhD agreement with the Universidad de La Laguna (ULL). S. Lemes-Perera acknowledges the funding and support provided by the company for this doctoral research.

The authors acknowledge support from the PID2022-138933OB-I00 ATQUE and 2023DIG28 IACTA research projects funded by MCIN/AEI/10.13039/501100011033/FEDER EU, and the CajaCanarias la Caixa Foundation, respectively. We also acknowledge the support provided by the CryptULL Research Group and the Cátedra de Ciberseguridad Binter-ULL for their contribution to the development of secure data management frameworks, as well as the institutional support provided by the Instituto Tecnológico y de Energías Renovables (ITER).

We express our gratitude to Light Bridges S.L. and ASTROPOC (\url{https://www.astropoc.com/}) for providing the high-performance GPU computing infrastructure, hardware resources, and astronomical datasets essential for the development, benchmarking, and performance evaluation of the \texttt{GpuFitsCrypt} library.


This article includes observations made in the Two-meter Twin Telescope (TTT) and Transient Survey Telescope (TST) both located at the Teide Observatory of the Instituto de Astrofísica de Canarias that Light Bridges operate on the island of Tenerife, Canary Islands (Spain). Dr. Antonio Maudes’s insights in economics and law were instrumental in shaping the development of this work.


\appendix
\section{Detailed Technical Resources}
\label{sec:appendix_start}

\subsection{Code Listings}
\label{app:code_listings}

This appendix provides the essential Python implementations for both the scientific integration workflow and the performance benchmarking protocol used in this study.

\subsubsection{Integration Workflow}
Listing~\ref{lst:python_example} illustrates how a scientist interacts with the library. It highlights the transparency of the solution: the complex GPU operations are abstracted, and the decrypted data is handed off to \texttt{Astropy} via zero-copy memory buffers.

\lstset{
    basicstyle=\footnotesize\ttfamily,
    columns=flexible,
    breaklines=true,
    captionpos=t,
    frame=tb,
    numbers=left,
    numberstyle=\tiny\color{gray},
    keywordstyle=\color{blue},
    commentstyle=\color{olive}\itshape,
    stringstyle=\color{purple},
    showstringspaces=false,
    xleftmargin=1.5em,
    framexleftmargin=1.5em
}

\begin{lstlisting}[language=Python, caption={Python integration workflow demonstrating granular access control. The code inspects the \texttt{warning\_code} to handle partial decryption scenarios (e.g., embargoed data) without crashing, allowing metadata analysis even when the pixel payload is suppressed.}, label={lst:python_example}]
import ctypes
from astropy.io import fits

# 1. Interface with the high-performance C++/CUDA backend
lib = ctypes.CDLL("./libgpufitscrypt.so")

# 2. Initialize Context (Pre-allocate GPU resources)
context = lib.gfc_context_create(MAX_FILE_SIZE, MAX_FILE_SIZE)

try:
    # 3. Decrypt and Authenticate (Parallel GPU execution)
    # Keys are typically retrieved from a KMS based on user role
    result = lib.gfc_decrypt_frame(context, ENCRYPTED_FILE, key_h, key_d)

    # 4. Granular Access Handling
    if result.error_code == 0:
        # Success: A valid FITS structure is available in memory
        fits_bytes = ctypes.string_at(result.data_buffer, result.buffer_size)
        
        with fits.HDUList.fromstring(fits_bytes) as hdul:
            # Check access level granted by the cryptographic engine
            if result.warning_code == 0:
                print("[FULL ACCESS] Processing scientific pixel data...")
                # ... perform photometry / source extraction ...

            elif result.warning_code == 2:
                # Scenario: User has Header Key but Data Key is invalid/embargoed
                print("[PARTIAL ACCESS] Payload Suppressed. Analyzing metadata only.")
                print("Target:", hdul[0].header['OBJECT']) # Metadata is valid
                # Pixel data is zero-filled; safe to read but contains no signal

            elif result.warning_code == 1:
                print("[FALLBACK] Header reconstruction. Original metadata unavailable.")

    else:
        # Fatal Error: File corruption or system failure
        print(f"[ERROR] Operation failed: {result.error_message}")

finally:
    lib.gfc_context_destroy(context)
\end{lstlisting}

\subsubsection{Benchmarking Methodology}
Listing~\ref{lst:benchmark_logic} details the experimental protocol described in Section~\ref{subsec:benchmarking}. Key features include the explicit separation of context initialization, the execution of discarded warm-up runs to stabilize the GPU state, and the isolation of the cryptographic overhead by subtracting the baseline I/O latency.

\lstset{
    basicstyle=\footnotesize\ttfamily,
    columns=flexible,
    breaklines=true,
    captionpos=t,
    frame=tb, 
    numbers=left,
    numberstyle=\tiny\color{gray},
    keywordstyle=\color{blue},
    commentstyle=\color{olive}\itshape,
    stringstyle=\color{purple},
    showstringspaces=false,
    xleftmargin=1.5em,
    framexleftmargin=1.5em
}

\begin{lstlisting}[language=Python, caption={Benchmarking logic demonstrating the experimental protocol: baseline I/O measurement, context reuse, warm-up cycles, and steady-state timing.}, label={lst:benchmark_logic}]
import time
import numpy as np
from astropy.io import fits

# Configuration constants
NUM_WARMUP_RUNS = 5
NUM_TIMED_RUNS = 200

def measure_baseline_io(original_file):
    """
    Measures the baseline I/O performance (Astropy Normal Open).
    """
    times = []
    for _ in range(NUM_TIMED_RUNS):
        start = time.monotonic()
        # memmap=False forces actual disk I/O, ensuring a fair
        # comparison against the decryption pipeline.
        with fits.open(original_file, mode='readonly', memmap=False) as hdul:
            _ = hdul[0].data.shape # Force header parsing
        times.append(time.monotonic() - start)
    return np.median(times)

def run_benchmark_session(lib, encrypted_file, key_h, key_d):
    """
    Executes the full benchmark protocol for a specific configuration.
    """
    # 1. Context Initialization (Once per session)
    # Allocates pinned memory and CUDA streams. Overhead excluded.
    context = lib.gfc_context_create(MAX_SIZE, MAX_SIZE)

    try:
        # 2. Warm-up Phase (Discarded)
        # Stabilizes GPU clocks and JIT compilation.
        for _ in range(NUM_WARMUP_RUNS):
            lib.gfc_decrypt_frame(context, encrypted_file, key_h, key_d)

        # 3. Steady-State Measurement Phase
        latencies = []
        for _ in range(NUM_TIMED_RUNS):
            t_start = time.monotonic()
            
            # Execute GPU Decryption & Authentication
            result = lib.gfc_decrypt_frame(context, encrypted_file, key_h, key_d)
            
            t_end = time.monotonic()
            
            if result.error_code == 0:
                latencies.append(t_end - t_start)

        return np.median(latencies), np.std(latencies)

    finally:
        # 4. Resource Teardown
        lib.gfc_context_destroy(context)

# Main Execution Flow
# -------------------
# A. Establish Baseline (I/O only)
baseline_io_time = measure_baseline_io("image_large.fits")

# B. Measure Cryptographic Performance (I/O + GPU + Overhead)
median_e2e, std_e2e = run_benchmark_session(lib, "enc_image_large.fits", ...)

# C. Calculate Net Cost
net_crypto_cost = median_e2e - baseline_io_time
print(f"Net Cryptographic Cost: {net_crypto_cost:.4f} s")
\end{lstlisting}


\subsection{Tables}
\label{app:tables_all}


\begin{table*}[t]
\centering
\caption{Detailed hardware specifications. Storage technology is a primary factor in I/O throughput bottlenecks during FITS processing.}
\label{tab:hardware_specs_detailed}
\begin{tabular}{l l l l l l}
\toprule
\textbf{Hostname} & \textbf{System Type} & \textbf{GPU Model (VRAM)} & \textbf{CPU Architecture} & \textbf{RAM} & \textbf{Storage Type} \\
\midrule
\texttt{AAZ} & HPC Node & NVIDIA H100 PCIe (80GB) & AMD EPYC 9124 & 62~GB & NVMe SSD \\
\texttt{ALE} & HPC Node & NVIDIA A100-SXM4 (80GB) & Intel Xeon Gold 6354 & 1~TB & NVMe SSD \\
\texttt{AHP} & HPC Node & NVIDIA L40S (48GB) & Intel Xeon Platinum 8468 & 251~GB & SAS SSD \\
\texttt{TTT} & Workstation & GeForce RTX 3090 (24GB) & Intel Core i9-10940X & 125~GB & SATA HDD (RAID) \\
\texttt{CCU} & Desktop & GeForce RTX 3060 (12GB) & Intel Core i5-10400F & 62~GB & NVMe SSD \\
\texttt{LLB} & Laptop & RTX 3050 Ti Mobile (4GB) & Intel Core i7-12700H & 31~GB & NVMe SSD \\
\texttt{JET} & Edge Device & NVIDIA Jetson Orin AGX & ARMv8 (Cortex-A78AE) & 8~GB & NVMe/eMMC \\
\bottomrule
\end{tabular}
\end{table*}


\subsection{Detailed Benchmark Configuration}
\label{app:benchmarks}

To ensure reproducibility, we detail the hardware and software configuration used for all performance evaluations.

\paragraph{Synthetic Dataset Generation}

Benchmarks were conducted using a suite of real and synthetic FITS images. Synthetic images were generated via \texttt{Astropy} to mimic real scientific observations where specific sizes were needed. The pixel data consists of 32-bit floating-point numbers (\texttt{BITPIX = -32}) drawn from a Gaussian distribution ($\mu=1500, \sigma=200$). The headers include standard WCS telemetry, observation metadata and flux calibration. Table~\ref{tab:dataset_specs} details the dimensions and sizes of the test files.


\begin{table}[t]
\centering
\caption{Synthetic FITS datasets for performance profiling. Sizes correspond to the pixel array payload (float32).}
\label{tab:dataset_specs}
\begin{tabular}{l l c r}
\toprule
\textbf{Category} & \textbf{Dimensions (px)} & \textbf{Pixels ($10^6$)} & \textbf{Size} \\
\midrule
Small       & $2048 \times 2048$     & 4.2   & 17~MB \\
Medium      & $7241 \times 7241$     & 52.4  & 201~MB \\
Large       & $14200 \times 10650$   & 151.2 & 577~MB \\
Extra Large & $30000 \times 30000$   & 900.0 & 3.35~GB \\
\bottomrule
\end{tabular}
\end{table}

\paragraph{Hardware Environment}
Benchmarks were executed on a diverse array of physical hosts, ranging from edge devices to HPC nodes. To ensure a consistent software environment, all tests were run within Docker containers based on \texttt{nvidia/cuda:12.0.1-devel-ubuntu22.04} (x86\_64) and \texttt{nvcr.io/nvidia/l4t-cuda:12.2.12-devel} (ARM64/Jetson), ensuring isolation and reproducibility. The detailed specifications are provided in Table~\ref{tab:hardware_specs_detailed} (see above).


\bibliographystyle{elsarticle-harv}
\bibliography{references_zotero_bibtex}

\end{document}